\documentclass[journal]{IEEEtran}

\usepackage{xcolor,soul,framed} 

\colorlet{shadecolor}{yellow}
\usepackage[pdftex]{graphicx}

\graphicspath{{../pdf/}{../jpeg/}}
\DeclareGraphicsExtensions{.pdf,.jpeg,.png}

\usepackage[cmex10]{amsmath}
\usepackage{array}
\usepackage{mdwmath}
\usepackage{mdwtab}
\usepackage{eqparbox}
\usepackage{url}

\usepackage{cite}
\usepackage{booktabs} 
\usepackage{multirow}
\usepackage{multicol}
\usepackage{amsfonts}
\usepackage{float}

\usepackage{amssymb}
\usepackage{pifont}
\newcommand{\cmark}{\ding{51}}%
\newcommand{\xmark}{\ding{55}}%

\begin{document}
    \title{Selective Listening by Synchronizing \\ Speech with Lips}
    \author{Zexu~Pan,
      Ruijie~Tao,
      Chenglin~Xu,~\IEEEmembership{Member,~IEEE,}
      and~Haizhou~Li,~\IEEEmembership{Fellow,~IEEE}

    \thanks{
    This research is supported by the Agency for Science, Technology and Research (A*STAR) under its AME Programmatic Funding Scheme (Project No. A18A2b0046); 
    The Science and Engineering Research Council, A*STAR, Singapore, through the National Robotics Program under Human-Robot Interaction Phase 1 (Grant No. 192 25 00054). 
    This research is also funded by the Deutsche Forschungsgemeinschaft (DFG, German Research Foundation) under Germany's Excellence Strategy (University Allowance, EXC 2077, University of Bremen), and by The Chinese University of Hong Kong, Shenzhen (UDF01002333, UF02002333). (\textit{Corresponding author: Chenglin Xu}).}
    \thanks{Zexu Pan is with the Integrative Sciences and Engineering Programme, and the Institute of Data Science, National University of Singapore, 119077 Singapore (e-mail: pan\_zexu@u.nus.edu).}
    \thanks{Zexu Pan, Ruijie Tao,  Chenglin Xu, and Haizhou Li are with the Department of Electrical and Computer Engineering, National University of Singapore, 119077 Singapore (e-mail: ruijie.tao@u.nus.edu; xuchenglin28@gmail.com; haizhou.li@u.nus.edu).}
    \thanks{Chenglin Xu is also with Kuaishou Technology, 518063 Shenzhen, China.}
    \thanks{Haizhou Li is also with The Chinese University of Hong Kong, Shenzhen, China, and the University of Bremen, 28359 Bremen, Germany.}
    }


\maketitle

\begin{abstract}
A speaker extraction algorithm seeks to extract the speech of a target speaker from a multi-talker speech mixture when given a cue that represents the target speaker, such as a pre-enrolled speech utterance, or an accompanying video track. Visual cues are particularly useful when a pre-enrolled speech is not available. In this work, we don't rely on the target speaker's pre-enrolled speech, but rather use the target speaker's face track as the speaker cue, that is referred to as the auxiliary reference, to form an attractor towards the target speaker. We advocate that the temporal synchronization between the speech and its accompanying lip movements is a direct and dominant audio-visual cue. Therefore, we propose a self-supervised pre-training strategy, to exploit the speech-lip synchronization cue for target speaker extraction, which allows us to leverage abundant unlabeled in-domain data. We transfer the knowledge from the pre-trained model to the attractor encoder of the speaker extraction network. We show that the proposed speaker extraction network outperforms various competitive baselines in terms of signal quality, perceptual quality, and intelligibility, achieving state-of-the-art performance.

\end{abstract}


\IEEEpeerreviewmaketitle

\section{Introduction}
\label{sec:introduction}
\IEEEPARstart{H}{umans} have a remarkable ability to focus attention on a particular speech signal in the presence of multiple noise sources and competing background speakers~\cite{cherry1953some}. The speaker extraction algorithm mimics human selective attention to extract only the target speaker's speech in such an adverse acoustic environment, which is also referred to as the cocktail party problem~\cite{bronkhorst2000cocktail}.

The human brain has a limited capacity to process all sensory stimuli perceived. Instead, human cognition relies on an inherent attention mechanism to focus the neural resources according to the contingencies of the moment~\cite{katsuki2014bottom}. Such attention can be categorized into a top-down and bottom-up process. Top-down attention refers to the voluntary allocation of neural resources based on guidance such as a specific goal or a  reference. Bottom-up attention is driven by external stimuli that are salient because of their inherent properties relative to the background~\cite{pinto2013bottom}. In this paper, we aim to mimic human visual top-down attention during listening to solve the single-channel target speaker extraction problem. This is a non-trivial, but crucial step for other downstream tasks such as speaker recognition~\cite{tao2020audio,Rao2019,xu2021target,liu2022neural,ma2021pl}, automatic speech recognition~\cite{yue2019end}, sound source localization~\cite{qian2021multi}, and voice conversion~\cite{zhou2020multi,zhou2021language,zhou2021seen}.

Humans separate a speech mixture by attending to the salient prosody of individual speech tracks. Speech separation represents one direction of research towards machine intelligence of such bottom-up auditory attention, which recently has seen major progress~\cite{hershey2016deep,luo2019conv,xu2018single,luo2020dual,chen2017deep,zeghidour2020wavesplit,stoller2018wave,kolbaek2017multitalker,liu2019divide}. The formulation of speech separation usually requires the number of speakers to be known in advance, which limits the scope of real-world applications. The formulation of speaker extraction avoids such limitation, that uses a reference signal to form the top-down attention, which is also referred to as the attractor in this paper. It proves to be effective in dealing with an unknown number of speakers. Most speaker extraction algorithms make use of auditory stimuli such as a reference speech signal to characterize the speakers. The reference speech signal is encoded as a speaker embedding in advance. In this way, the speaker extraction algorithm extracts speech that sounds similar to the reference speech signal~\cite{Chenglin2020spex,spex_plus2020,wang2019voicefilter,he2020speakerfilter,8462661,vzmolikova2019speakerbeam,xiao2019single,shi2020speaker,delcroix2020improving,sato2021multimodal,ochiai2019multimodal,marvin2021,wang21aa_interspeech}.

As the above-mentioned speaker extraction techniques use the speaker embedding as the attractor in the neural solutions, they require pre-enrollment of a reference speech signal, which sometimes is not available in practice, for example, when we would like a robot to pay auditory attention to a passer-by. Unlike the reference speech signal, the real-time video recording can be easily obtained during human-computer interactions. In this work, we seek to develop a speaker extraction solution, which doesn't require a pre-enrolled speech signal as the reference, but assumes that the target speaker's face is visually present. We use the target speaker's face track as the auxiliary reference for the target speaker extraction.

Human attention is multi-modal~\cite{smith2005development}, through a variety of sensory systems such as audition, vision, touch, and smell. These multi-modal stimuli are processed in human nervous systems in an interactive manner, which is referred to as the \textit{reentry}~\cite{edelman1987neural}. The \textit{reentry} theory describes the bidirectional exchange of signals along reciprocal axonal fibers linking two or more brain areas, that are temporally correlated, and can educate each other. The human brain has a cortex to process the audio and visual stimuli together~\cite{fleming2020audio}. In a cocktail party, visual cues are not corrupted by background noise, reverberation, and interference speech thus providing a robust attractor for auditory attention~\cite{michelsanti2021overview}. The studies in neuroscience show that speech comprehension is improved by looking at the speaker in adverse conditions~\cite{golumbic2013visual,crosse2016eye}, leveraging on the long-term cross-modal temporal integration. Inspired by these findings, we aim to emulate the human visual top-down attention in a computational model for speaker extraction, named the \textit{reentry} model after the \textit{reentry} theory by Edelman~\cite{edelman1987neural}. 

In human selective listening, there are typically three types of visual cues, namely face-voice association~\cite{chung2020facefilter,qu2020multimodal,gao2021visualvoice}, viseme-phoneme mapping~\cite{afouras2018conversation,wu2019time,nguyen2020deep,li2020deep,pan2021usev}, and speech-lip synchronization~\cite{lu2019audio,gabbay2017visual,morrone2019face,lu2018listen,sodoyer2004developing}. The face-voice association explores the general correlation between a human face and its voice signatures such as gender, age, and nationality. The viseme-phoneme mapping defines the relationship between the shapes of lips and mouth with respect to the sounds they represent, i.e., phonemes~\cite{bear2017phoneme}.  Among the visual cues, we advocate that the temporal synchronization between speech and lip movements would be the most informative, which will be the focus of this paper. The study on speech-lip temporal synchronization is a departure from the time-domain audio-visual speaker separation~\cite{wu2019time} and
the multi-modal speaker extraction~\cite{pan2020muse} model, where a supervised pre-training strategy is employed to explore the viseme-phoneme mapping cue for speaker extraction. 

To emulate human visual top-down attention during listening in the cocktail party scenario, we explore the speech-lip synchronization in a multi-talker setting with a pre-trained network, named the speech-lip synchronization (SLSyn) network. We propose a self-supervised training strategy for the SLSyn network such that the learning of speech-lip synchronization could leverage abundant unlabeled training data that are in the same domain as the speaker extraction task. We transfer the pre-trained knowledge from the SLSyn network to the \textit{reentry} model, which performs `listening with ears and eyes'. The proposed \textit{reentry} model outperforms various competitive baselines and achieves the state-of-the-art performance on VoxCeleb2~\cite{Chung18b} dataset mixtures, in terms of signal quality, perceptual quality, and intelligibility. The \textit{reentry} model also shows its robustness in cross-dataset evaluations and visual occlusion evaluations.

The rest of the paper is organized as follows. In Section~\ref{sec:tse}, we introduce the related works on target speaker extraction. In Section~\ref{sec:reentry}, we formulate the proposed \textit{reentry} model. In Section~\ref{sec:experimental_setup}, we describe the experimental setup. In Section~\ref{sec:results}, we report the results. Finally, Section~\ref{sec:conclusion} concludes the study.

\begin{figure*}
  \centering
  \includegraphics[width=0.99\linewidth]{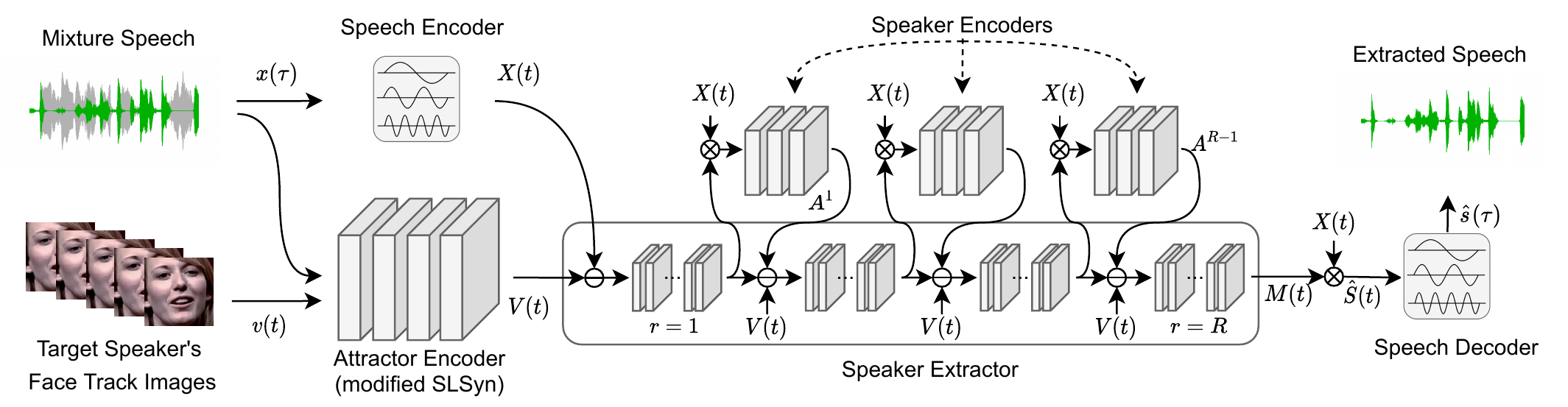}
  \caption{The proposed audio-visual speaker extraction network, named the \textit{reentry} model. The attractor encoder derives feature representations for speech-lip synchronization; the speaker extractor produces a receptive mask for the target speaker dynamically; the speaker encoders characterize the target speaker and provide the top-down attention to the speaker extractor. The symbol $\ominus$ represents the concatenation of features along channel dimension in convolutional layers; the symbol $\otimes$ represents element-wise multiplication.}
  \label{fig:reentry}
\end{figure*}

\section{Reference Signals for Speaker Extraction}
\label{sec:tse}
Neural speaker extraction typically requires a reference signal to form the top-down attention towards the voice of the target speaker in a speech mixture. Studies in neuroscience suggest that either auditory~\cite{hill2010auditory} or visual stimulus~\cite{golumbic2013visual,crosse2016eye} can serve as a reference cue in human selective listening.

\subsection{Pre-enrolled auditory stimuli}
Each speaker has a unique voice signature, which can be characterized by a fixed dimensional vector, referred to as speaker embedding, such as i-vector~\cite{dehak2010front}, x-vector~\cite{snyder2016deep}, d-vector~\cite{wan2018generalized}, and other similar feature representations~\cite{huang2018angular}. 

It is common that a speaker embedding derived from the target speaker is employed to form the attractor for speaker extraction. VoiceFilter~\cite{wang2019voicefilter} and TseNet~\cite{9004016} are such examples, where they pre-train a speaker encoder, which is then used to derive d-vector or i-vector from the reference speech signal, to extract the target speaker.
SpEx~\cite{Chenglin2020spex} and SpEx+~\cite{spex_plus2020} further the idea by introducing a multi-task learning strategy, to jointly train the speaker encoder with the speaker extractor network, thus benefiting from the improved speaker characterization~\cite{wang2019voicefilter,9004016} as well as the task-oriented optimization~\cite{8462661,8268910,8683874}. 

\subsection{Self-enrolled auditory stimuli}
Despite much success, neural speaker extraction with speaker pre-enrollment is not practical in some scenarios. This prompts us to look into solutions without the need for a pre-enrolled reference speech signal. 
A recent study~\cite{afouras2019my} on audio-visual speaker extraction seeks to use a visual reference to replace an audio reference. In this study, the visual reference is employed to extract the target speech first, which is subsequently encoded as a speaker embedding to serve as the reference whenever the visual cues are occluded. While such a two-pass mechanism improves in the absence of visual cues, the self-enrolled speaker embedding does not contribute when visual cues are adequately available. This suggests that the self-enrolled speaker embedding is not as informative as the visual reference, partly due to the fact that the two-pass mechanism is not jointly optimized.

MuSE~\cite{pan2020muse} furthers the idea of using a visual reference in audio-visual speaker extraction. It self-enrolls the speaker's voice signature from the intermediate extracted target speech on the fly. MuSE leverages the visual reference as well as the self-enrolled speaker embedding in an integrated manner, which shows that both of them actively contribute to the speaker extraction task. AVSE~\cite{9414133} is another method that self-enrolls the speaker's voice signature directly from the speech mixture. AVSE is a frequency-domain implementation while MuSE takes a time-domain approach. This paper is motivated by the findings in the prior studies on the effective use of visual reference to self-enroll a speaker.

\subsection{Face-voice association}
There exist general correlations between the human face and voice.
Prior studies~\cite{chung2020facefilter,qu2020multimodal} have explored such knowledge for speaker extraction. They employ a speaker encoder that is pre-trained to learn the cross-modal correspondence and to encode a face image into a speaker embedding as the attractor. 
It is noted that speaker extraction systems that rely on such general correlation knowledge generally don't outperform those with the pre-enrolled auditory stimuli.

\subsection{Viseme-phoneme mapping}
Phonemes are the smallest units of spoken language, and the visual equivalent are the visemes. Generally speaking, a viseme corresponds to a set of phonemes that have identical appearances on the lips. Therefore, one is able to associate a phoneme with a viseme class, but a viseme may map to multiple phonemes. The viseme-phoneme mapping is usually trained from a visual speech recognition task, which aligns the input viseme sequence with the output phoneme sequence or word sequence~\cite{Afouras18b,Chung16,lu2021visualtts}.

Some prior studies have effectively made use of such viseme-phoneme mapping in speaker extraction, i.e. TDSE~\cite{wu2019time} and MuSE~\cite{pan2020muse}. They transfer part of the pre-trained visual speech recognition network as the visual encoder of the speaker extraction network, where the visual encoder encodes the input viseme sequence to a viseme embedding sequence. The viseme embeddings are used to form the attention attractor. As the visual encoder is optimized for the phoneme recognition task, therefore, not the best for speech reconstruction, we are prompted to look into a better way to train the visual encoder for speaker extraction. Furthermore, the training of visual speech recognizer requires video data with text transcription, which is scarce. We are motivated to leverage the audio-visual information that exists in abundantly available parallel audio-visual data for self-supervised learning, without the need for text transcription.

Motivated by a study that face embeddings encoded by face recognition models can recover facial expressions~\cite{rudd2016moon}, `looking to listen at the cocktail party'~\cite{ephrat2018looking} employs viseme embeddings which are encoded by a face recognition model as the attractor for speaker extraction. However, the face embeddings are trained to discriminate faces. It is neither an optimal representation for viseme-phoneme mapping nor audio-visual synchronization. 

\subsection{Audio-visual synchronization}  
Some studies have implemented the use of audio-visual synchronization cue for speaker extraction~\cite{gabbay2017visual,morrone2019face}, where they encode the raw visual frames or facial landmarks into a sequence of visual embeddings as the reference signal.

Others~\cite{owens2018audio, afouras2020self} further the idea by pre-training the visual encoder of the speech separation network on audio-visual synchronization tasks. \cite{owens2018audio} pre-trains the visual encoder on the scene synchronization task which finds that the scene motion is temporally synchronized with audio events. \cite{afouras2020self} pre-trains the visual encoder on the object synchronization task which detects the object that is responsible for the sound in a video. All studies point to the direction that the scene and object synchronization features encoded by the visual encoder could be employed as a visual attractor in speech separation networks. However, the audio-visual synchronization for speech to lips in a multi-talker scenario is not as prominent as that for whistling sound to a train. Furthermore, the audio-object synchronization features are not pre-trained to disentangle overlapped speech.

The success of the effective use of the audio-visual synchronization cue prompts us to look into speech-lip synchronization that is a more direct and informative cue for speaker extraction. Humans have the experience to improve listening by observing the lips when someone is talking. In this paper, we would like to focus on the use of the speech-lip synchronization cue for speaker extraction. 

\begin{figure*}
    \begin{minipage}[b]{.64\linewidth}
      \centering
      \centerline{\includegraphics[height=5cm]{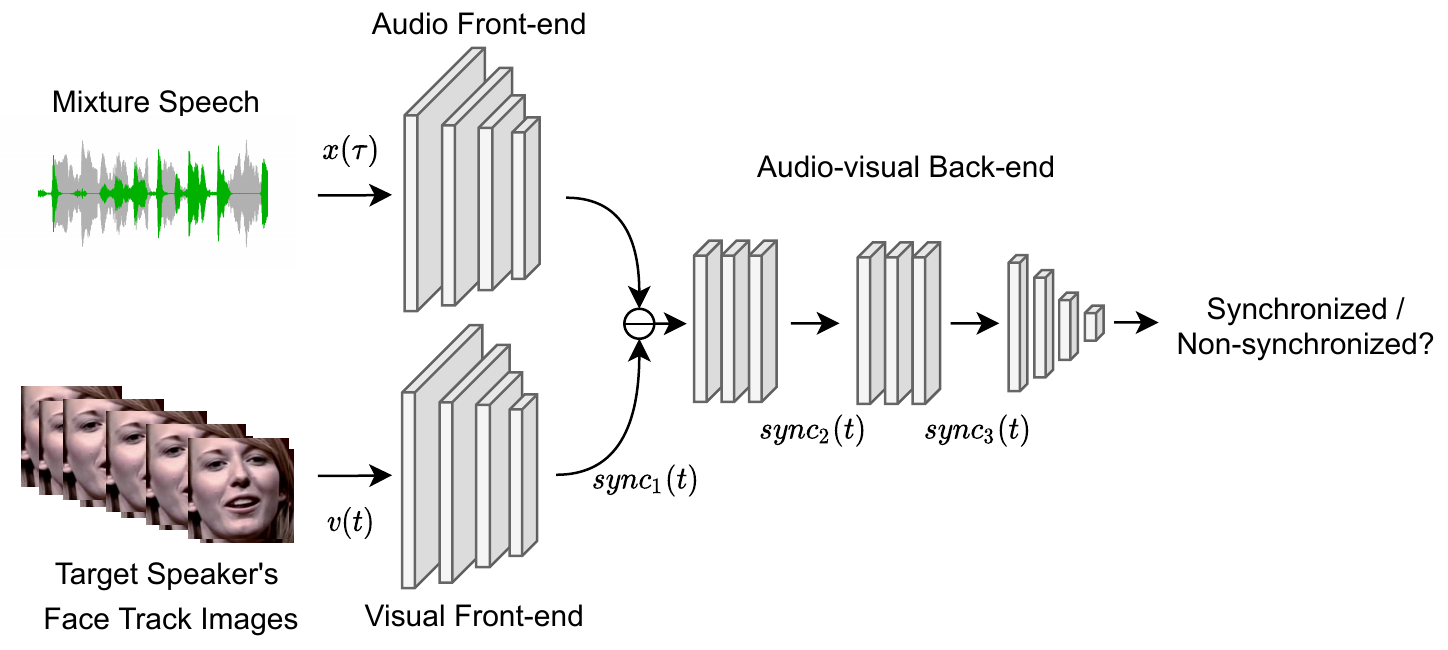}}
      \centerline{\scalebox{0.8}{(a) The speech-lip synchronization (SLSyn) network.}}\medskip
    \end{minipage}
    \hfill
    \begin{minipage}[b]{0.35\linewidth}
      \centering
      \centerline{\includegraphics[height=5cm]{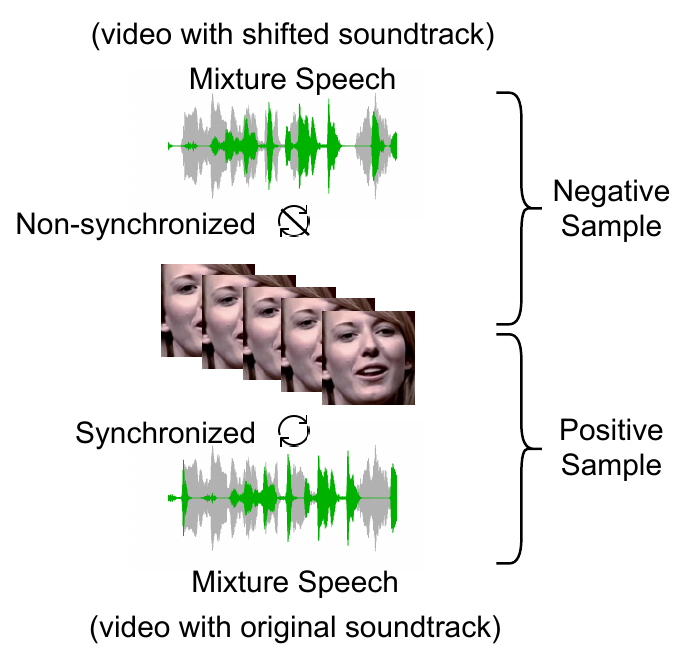}}
      \centerline{\scalebox{0.8}{(b) Positive/negative samples} }\medskip
    \end{minipage}
    \caption{The proposed speaker-independent speech-lip synchronization network, that learns to extract speech-lip synchronization embeddings. In (a), the SLSyn network takes the face track and the soundtrack as inputs and makes a synchronization decision. Part of the pre-trained SLSyn network is used as the attractor encoder in Fig.~\ref{fig:reentry}. In (b), we illustrate a positive (synchronized) and a negative (non-synchronized) audio-visual sample. The green waveform for a target speaker in the synchronized sample is shifted to form a negative sample.}
    \label{fig:slsynet}
\end{figure*}

\section{Audio-visual speaker extraction network}
\label{sec:reentry}
We propose an audio-visual speaker extraction network using the speech-lip synchronization cue, as depicted in Fig.~\ref{fig:reentry}, that is called the \textit{reentry} model. Let $x(\tau)$ be a multi-talker speech waveform in the time-domain, which consists of the target speaker's speech $s(\tau)$ and other interference speech $b_{i}(\tau)$:
\begin{equation}
    \label{eqa:speaker_extraction}
    x(\tau) = s(\tau) + \sum_{i=1}^{I}b_{i}(\tau)
\end{equation}
We would like to extract $\hat{s}(\tau)$ that approximates $s(\tau)$.

In the \textit{reentry} model: 1) The speech encoder transforms $x(\tau)$ into a sequence of spectrum-like frame-based embeddings $X(t)$, referred to as the mixture speech embeddings\footnote{In this paper, a variable with $\tau$ as the index represents a sequence of samples in the time-domain, while a variable with $t$ as the index represents a sequence of frame-based embeddings.}. 2) The attractor encoder encodes $x(\tau)$ and its accompanying face track images $v(t)$ into a sequence of speech-lip synchronization embeddings $V(t)$ to form an attractor signal. 3) The speaker encoders encode the intermediate estimated speech embeddings into a speaker embedding $A^r$, which represents the attended speaker's voice signature. 4) The speaker extractor takes in $V(t)$ and $A^r$ as top-down attention attractors to estimate a mask $M(t)$ for $X(t)$ which only lets the target speaker pass. The extracted speech embeddings $\hat{S}(t)$ are estimated by element-wise multiplication between the estimated mask $M(t)$ and the mixture speech embeddings $X(t)$:
\begin{equation}
    \hat{S}(t) = X(t) \otimes M(t)
\end{equation}
5) Finally, the speech decoder transforms the extracted speech embeddings $\hat{S}(t)$ into a time-domain waveform $\hat{s}(\tau)$.

\subsection{Self-supervised speech-lip synchronization learning}
\label{sec:slsynet}
Traditional approaches have been devoted to using statistical theory to learn audio-visual synchronization~\cite{fisher2004speaker,hershey1999audio}. With deep learning, audio-visual synchronization learning has seen major progress with self-supervised learning, where a network learns from positive and negative samples easily derived from found data. The positive samples could be the videos with the original soundtrack, while the negative samples are the videos with a random time-shifted soundtrack~\cite{owens2018audio,afouras2020self,marcheret2015detecting}.

During learning, one common approach is to encode audio and visual signals into embedding sequences separately, and compare the similarity of the two embedding sequences~\cite{chung2016out,chung2019perfect,afouras2020self}. It makes the decision on short video segments, and doesn't make good use of the temporal correlation between modalities. Another idea is to fuse the audio and visual embedding sequences, that are extracted separately, through an audio-visual network such as a convolutional neural network~\cite{owens2018audio} or a long short-term memory (LSTM) network~\cite{shalev2020end} to capture their temporal dependencies. 

Motivated by the second idea, we propose a speech-lip synchronization (SLSyn) network to detect the presence of synchronization between a multi-talker speech signal and a face track. As the SLSyn network is optimized for speech-lip synchronization detection, it is expected to derive speech-lip synchronization embeddings that characterize the audio-visual joint events. The SLSyn network is pre-trained with multiple speakers to be a speaker-independent model. 
 
As shown in Fig.~\ref{fig:slsynet}(a), the SLSyn network consists of an audio front-end to extract audio features from the mixture speech $x(\tau)$, a visual front-end to extract visual features from the target speaker's face track images $v(t)$, and an audio-visual back-end to fuse the concatenated audio and visual features for an utterance level binary classification. As the temporal convolutional neural network (TCN) is effective in capturing long-term dependencies~\cite{luo2019conv,tao2021someone} in speech processing~\cite{oord2016wavenet,rethage2018wavenet}, we adopt a TCN structure as the audio front-end. We design a visual front-end with a 3D convolutional layer and a ResNet to encode raw images into visemes similar to~\cite{stafylakis2017combining}, representing the appearance of the lips. The audio and visual features are concatenated at the frame level, and taken by the TCN-based audio-visual back-end.

The activations at the intermediate layers of the SLSyn network represent different levels of abstraction. We study three different feature representations as the speech-lip synchronization embeddings, namely $sync_{1}(t)$, $sync_{2}(t)$, and $sync_{3}(t)$, as shown in Fig.~\ref{fig:slsynet}(a). The three feature representations have the same time-resolution as the target speaker's face track.

As illustrated in Fig.~\ref{fig:slsynet}(b), the original videos with speech-lip synchronized soundtracks serve as the positive samples, while videos with randomly shifted soundtracks serve as the negative samples. An additional interference speech signal is added to both the positive and negative samples to simulate the cocktail party scenario. 

We minimize the following binary cross-entropy (BCE) loss for the SLSyn network training:
\begin{equation}
    \mathcal{L}_{BCE} = - ylog(\hat{y}) - (1-y)log(1-\hat{y})
\end{equation}
where $y \in \{0,1\}$ indicates whether the input video has a shifted soundtrack while $\hat{y}$ is the predicted probability of the video and soundtrack being synchronized at the end of the SLSyn network. 

This learning is similar to the noise-contrastive estimation~\cite{gutmann2010noise}, where the noisy samples in the SLSyn network training are asynchronous videos. As human-annotated labels are not needed, the pre-training can be done on abundantly available videos, also in the same domain as the speaker extraction algorithms to minimize the domain mismatch. 

\subsection{Speaker extraction using the speech-lip synchronization cue}
We now formulate the \textit{reentry} model, as illustrated in Fig.~\ref{fig:reentry}. The \textit{reentry} model consists of a speech encoder, an attractor encoder, a speaker extractor,  a number of speaker encoders, and a speech decoder. The speaker extractor has an interlace structure between a number of speaker encoders and speaker extraction modules. 

\subsubsection{Speech encoder}
\label{sec:speech_encoder}
The speech encoder takes a time-domain approach~\cite{luo2019conv} to encode the input speech waveform $x(\tau)$ into a spectrum-like speech embedding sequence $X(t)$. It consists of a 1D convolution $Conv1D$, with channel size $N$, kernel size $L$, and stride $L/2$. The $Conv1D$ is followed by a rectified linear activation $ReLU$,

\begin{equation}
    \label{eqa:encoder}
    X(t) = ReLU(Conv1D(x(\tau), N, L, L/2))
\end{equation}
where $N$ is also the speech embedding dimension. As $x(\tau)$ is a time-domain signal, the 1D convolution behaves like a frequency analyzer~\cite{luo2019conv}. 

\subsubsection{Attractor encoder}
\label{sec:visual_encoder}
The attractor encoder seeks to encode the synchronized sequence of images $v(t)$ with the mixture speech waveform $x(\tau)$ into a sequence of embeddings $V(t)$ representing the speech-lip synchronization. The SLSyn network is trained just to do that. 

We adopt part of the pre-trained SLSyn network, followed by several adaptation layers to form the attractor encoder. The speech-lip synchronization embeddings extracted from the pre-trained SLSyn network, namely $sync_{1}(t)$, $sync_{2}(t)$, and $sync_{3}(t)$, are not optimized for speaker extraction directly, therefore, they have different properties than those from the speaker extractor~\cite{pan2020multi}. The adaptation layers adapt the speech-lip synchronization embeddings towards the speaker extraction task. We follow~\cite{afouras2018conversation,wu2019time} to design a stack of TCNs in the adaptation layers.

\subsubsection{Speaker extractor}
\label{sec:speaker_extractor}
Receptive masks have been well studied in the source separation literature to mimic human selective attention, such as ideal binary mask~\cite{li2009optimality}, ideal ratio mask~\cite{narayanan2013ideal}, ideal amplitude mask~\cite{wang2014training}, Wiener-filter like mask~\cite{erdogan2015phase}, and phase sensitive mask~\cite{erdogan2015phase}. We adopt the masking methods~\cite{spex_plus2020,luo2019conv,wu2019time} to estimate a receptive mask $M(t)$ for the speech of the target speaker as shown Fig.~\ref{fig:reentry}.

Similar to~\cite{Chenglin2020spex,luo2019conv,wu2019time}, we use the TCN structure for the speaker extractor to capture the long-range dependency of the speech signal. The speaker extractor consists of $R$ repeated stacks of TCNs in a speaker extraction pipeline. We illustrate the reentry model with $R=4$ in Fig.~\ref{fig:reentry}. An intermediate mask $M^r(t)$ is estimated after the $r^{th}$ stack. Each stack consists of $B$ TCNs with an exponentially growing dilation factor $2^b$ ($b = 0, ..., B-1$). With the pipelined stacks, the receptive field of the network increases exponentially, and better models the long-term temporal dependencies of speech signals~\cite{luo2019conv}. Hence, the masks produced by the TCN stacks are expected to be progressively refined along the speaker extraction pipeline.

The \textit{reentry} theory suggests that in the human brain, temporally correlated signals educate each other through bi-directional exchanges. Following the \textit{reentry} theory, we propose to make use of the interactive evidence between the speech and lips on the fly as we decode the target speech.

The speaker extractor takes the mixture speech embeddings $X(t)$ as input, and the top-down attention cues as the condition, to extract the clean speech from the target speaker. Besides the speech-lip synchronization cue $V(t)$, we also consider the self-enrolled speaker embedding $A^r$ as another top-down attention cue. The top-down attention cues modulate with the input mixture speech to estimate the receptive masks. The study in SpEx~\cite{Chenglin2020spex} shows that concatenating the attention stimulus at the start of every TCN stack effectively guides the mask estimation. Bringing the idea forward to the audio-visual context, we concatenate $A^r$ and $V(t)$ with $M^r(t)$ to form the input of each TCN stack, except for the first TCN stack which we concatenate $V(t)$ with $X(t)$ as input. The $A^r$ and $V(t)$ are up-sampled along the time dimension to match the temporal resolution of $M^r(t)$ before concatenation. We will discuss the self-enrolled speaker encoders next.

\subsubsection{Speaker encoders}
\label{sec:speaker_encoder}
A speaker encoder derives an utterance level speaker embedding $A^r$ a.k.a self-enrolled speaker embedding from the intermediate estimated speech embeddings $\hat{S}^r(t)$ after the $r^{th}$ TCN stack in the speaker extraction pipeline. The $\hat{S}^r(t)$ is defined as:
\begin{equation}
    \hat{S}^r(t) = X(t) \otimes M^r(t)
\end{equation}

Since $M^r(t)$ becomes available only after the first TCN stack, we apply the speaker encoders starting from the second TCN stack. We concatenate $A^r$ with $M^r(t)$ in the same way as $V(t)$. As $\hat{S}^r(t)$ varies along the speaker extraction pipeline, we design $R-1$ speaker encoders of the same architecture, but with distinctive weights that are trained independently of each other, as opposed to weight sharing in SpEx~\cite{Chenglin2020spex}. 

\subsubsection{Speech decoder}
\label{sec:speech_decoder}
The speech decoder is an inverse function of the speech encoder, that takes the extracted speech embeddings $\hat{S}(t)$ as input and reconstructs the time-domain waveform $\hat{s}(\tau)$ as output. We formulate the speech decoder as a linear layer followed by an overlap and add $OnA$ operation~\cite{oppenheim1978theory}:
\begin{equation}
    \label{eqa: decoder}
    \hat{s}(\tau) = OnA(Linear(\hat{S}(t), N, L), L/2)
\end{equation}
where the $Linear$ layer has input and output feature dimension of $N$ and $L$ respectively, and the overlap and add operation has a frame shift of $L/2$.

\subsection{Multi-task learning}
\label{sec:multi-task}
To ensure that the speaker embedding $A^r$ is encoded for the purpose of speaker extraction, we train the $R-1$ speaker encoders together with the speaker extractor network, with two training objectives, i.e., the scale-invariant signal-to-noise ratio (SI-SDR) loss~\cite{le2019sdr} to measure the signal quality, and the cross-entropy (CE) loss for speaker classification. 

The loss $\mathcal{L}_{SI\mbox{-}SDR}$ measuring signal quality is shown in Eq.~\ref{eqa:loss_sisnr}, which is applied to the output of the \textit{reentry} model between the extracted speech $\hat{s}(\tau)$ and clean speech $s(\tau)$. We omit the subscript ($\tau$) for brevity:
\begin{equation}
    \label{eqa:loss_sisnr}
    \mathcal{L}_{SI\mbox{-}SDR} = - 10 \log_{10} ( \frac{||\frac{<\hat{s},s>s}{||s||^2}||^2}{||\hat{s} - \frac{<\hat{s},s>s}{||s||^2}||^2})
\end{equation}

The loss $\mathcal{L}_{CE}^r$ that accounts for speaker classification is shown in Eq.~\ref{eqa:loss_ce}, which is applied to every speaker encoder's output $A^r$ of the $R-1$ speaker encoders:

\begin{equation}
    \label{eqa:loss_ce}
    \mathcal{L}_{CE}^r = - \sum_{c=1}^{C} p[c] \log \hat{p}^r[c]
\end{equation}
where $p[c]=1$ for target speaker, and $p[c]=0$ otherwise, $\hat{p}^r[c]=softmax(W^r A^r)[c]$ is the predicted probability for speaker $c$, $C$ is the number of speakers in the dataset, and $W^r$ is a learnable weight matrix specific to each speaker encoder for speaker classification. 

The overall loss $\mathcal{L}_{total}$ is defined as:
\begin{equation}
    \label{eqa:loss_all}
    \mathcal{L}_{total} = \mathcal{L}_{SI\mbox{-}SDR} + \gamma \sum_{r=1}^{R-1}\mathcal{L}_{CE}^r
\end{equation}
where $\gamma$ is a scaling factor. 

Now that each speaker encoder is constrained with a speaker classification loss, it is expected to derive a speaker embedding $A^r$ that best represents the target speaker's voice signature, helping the speaker extractor in the next TCN stack. Besides, the gradient of the speaker classification loss is passed from the speaker encoders to the stacks of TCNs in the multi-task learning framework. Hence, the speaker extractor is also trained to extract the target speaker's speech containing the target speaker's voice signature, besides the signal extraction quality. Therefore, the speaker self-enrollment is progressively refined with the speaker extractor with the interlaced structure, ensuring the convergence of speaker extraction.

It is worth noting that $A^r$ is derived from the mixture speech and visual signals without the supervision of speaker identity at run-time inference. Therefore, $A^r$ is also called the self-enrolled speaker embedding.

\subsection{Relationship with MuSE}

MuSE~\cite{pan2020muse} employs a pre-trained visual speech recognition network as the attractor encoder and exploits the viseme-phoneme mapping to form the attention attractor. The main difference between the \textit{reentry} model and MuSE lies in two aspects: 1) The \textit{reentry} model employs a self-supervised pre-training strategy for the SLSyn network that doesn't require text transcription of the soundtrack as the visual speech recognition network does. This brings us multiple advantages. We can now easily adapt the SLSyn network on abundantly available domain data to reduce domain mismatch. 2) The SLSyn network learns the speech-lip synchronization rather than the viseme-phoneme mapping. The latter is not a direct cue for overlapped speech disentanglement and signal reconstruction.
 
While the \textit{reentry} model is different from the MuSE framework, it shares some ideas that are studied in MuSE and other related work, that is to exploit a self-enrollment strategy, and to employ a multi-stage progressive refinement pipeline for speaker extraction. We expect that the progressive refinement architecture will outperform a single-stage extraction~\cite{ge2020multi}. The findings in this paper certainly reinforce this line of thought.

\section{Experimental Setup}
\label{sec:experimental_setup}
\subsection{Experiment data}
\subsubsection{Speaker extraction}
We simulate a two-speaker mixture dataset (VoxCeleb2-2mix) and a three-speaker mixture dataset (VoxCeleb2-3mix) from the VoxCeleb2 dataset~\cite{Chung18b} for a series of systems evaluations. The VoxCeleb2 dataset contains over 1 million face track videos for 6112 celebrities extracted from YouTube videos. The face track videos are obtained using a face detector as in~\cite{liu2016ssd}, and a tracker that is based on ROI overlap~\cite{Nagrani2017}. As the face track video comes with a soundtrack, each utterance is thus temporally synchronized with the speaker's face track.

To simulate the mixture datasets, we remove utterances that are shorter than $4$ seconds, and randomly select 40000 clean utterances from 800 speakers in the original VoxCeleb2 train set to create our train set (20000 utterances); we randomly select another 8000 clean utterances from the same 800 speakers in the original train set to create our validation set (5000 utterances);  we use the 36237 clean utterances from 118 speakers in the original test set to form our test set (3000 utterances). The interference speech is mixed with the target speech at a random Signal-to-Noise (SNR) ratio between $10$dB to $-10$dB. The long utterance is truncated to the length of the short utterance when creating a mixture speech. 
The sampling rate of the audio data is $16$ kHz, and the face track video stream is available at $25$ FPS. 
We use a sequence of sampled face track images from the target speaker's utterance as the visual input to our proposed network.

The train set is used for the network training, and the validation set is used for optimizing the network configurations. The utterances and speakers in the test set do not overlap with those in the train set and validation set, which allows us to perform speaker-independent experiments using the test set. 

\subsubsection{Synchronization detection}
To pre-train the speaker-independent speech-lip synchronization network, we create a multi-talker dataset (VoxCeleb2-sync) from the original VoxCeleb2 dataset~\cite{Chung18b}. We create $4$ million video clips to form a train set, and $40$ thousand video clips to form a validation set. They range from $1$ to $4$ seconds. A quarter of the video clips are with clean speech and three-quarters of them are with simulated two-speaker mixture speech. The simulation follows the same protocol as that of the VoxCeleb2-2mix dataset. 

The video clips with the original soundtrack are the positive samples that have synchronized speech and lip movements. The negative samples have a randomly shifted soundtrack by $0.2$ to $1$ seconds relative to the visual sequence. Besides the VoxCeleb2-sync dataset, we create a smaller version of the dataset named the VoxCeleb2-sync-s, which contains $10\%$ of the video clips from the VoxCeleb2-sync dataset; and a clean dataset named the VoxCeleb2-sync-c, which is of the same size as the VoxCeleb2-sync, but all are clean speech utterances from the VoxCeleb2 dataset.

\subsection{Model configuration}

\begin{figure}
  \centering
  \includegraphics[width=0.8\linewidth]{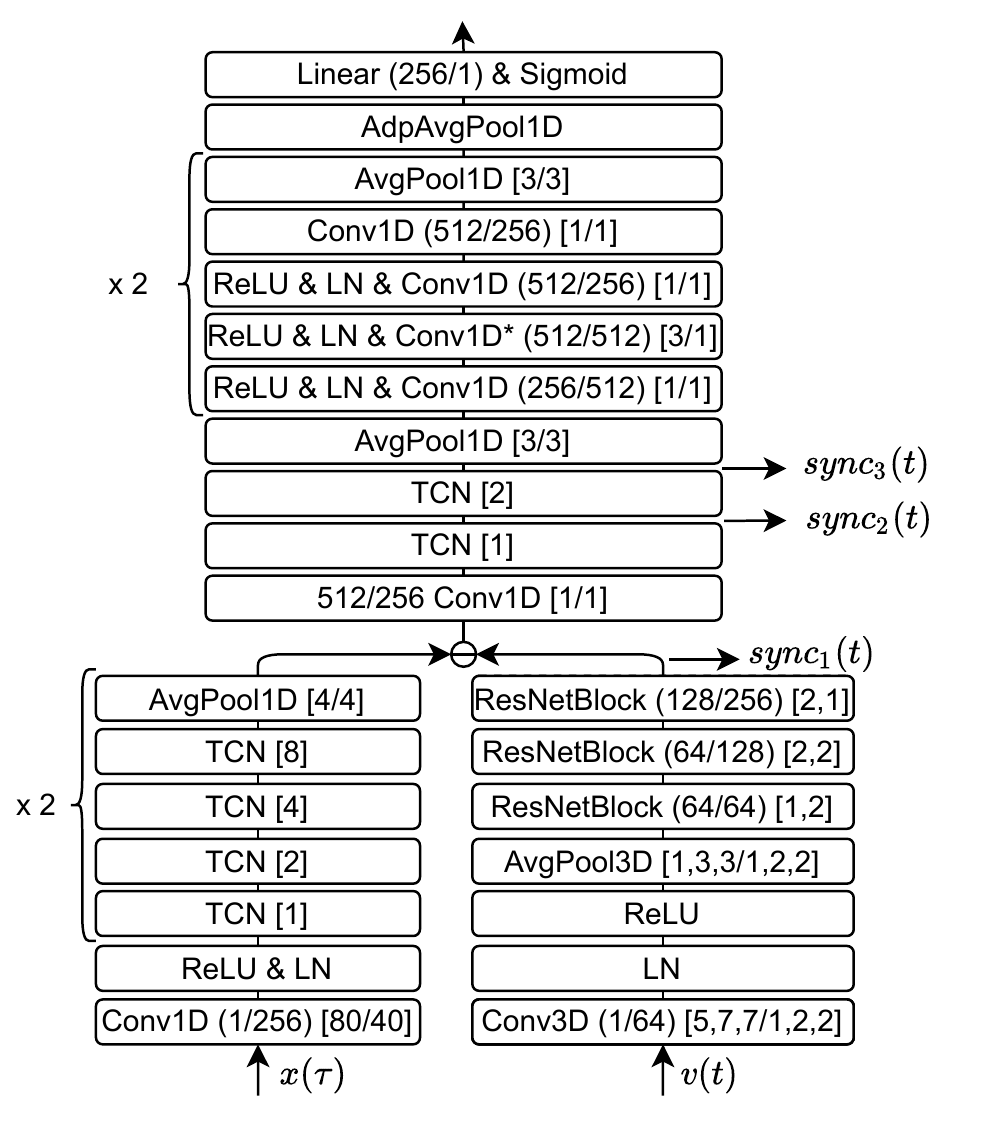}
  \caption{The configuration of the SLSyn network. Throughout the paper, $Conv1D$, $Conv2D$, and $Conv3D$ are 1, 2, and 3-dimensional convolutional layers respectively, and their parameters are represented as $(in\_channels/out\_channels) [kernel\_size/stride]$, $Conv1D^*$ is group convolution, with group size $in\_channels$, $LN$ is layer normalization, $AvgPool1D$ and $AvgPool3D$ are 1 and 3-dimensional average pooling layers respectively, and their parameters are represented as $[kernel\_size/stride]$, $AdpAvgPool1D$ is the adaptive 1-dimensional average pooling layer, and $PReLU$ is parametric $ReLU$. The configuration of the TCN and the ResNetBlock are detailed in Fig.~\ref{fig:tcn&mts_resnet}(a) and (b). }
  \label{fig:mts}
\end{figure}

The SLSyn network as shown in Fig.~\ref{fig:slsynet} consists of an audio front-end, a visual front-end, and an audio-visual back-end. The model configuration is detailed in Fig.~\ref{fig:mts}. The last linear layer of the SLSyn network has an input channel size of 256 and an output channel size of 1, a $Sigmoid$ activation is used at the end for binary classification. 

\begin{figure}
\begin{minipage}[b]{.25\linewidth}
  \centering
  \centerline{\includegraphics[width=\linewidth]{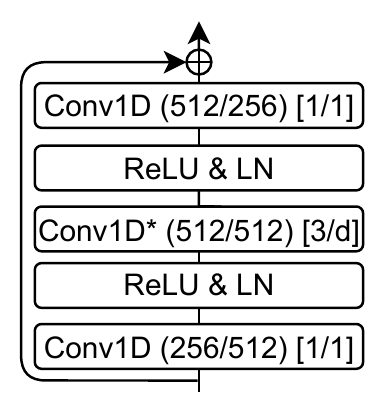}}
  \centerline{\scalebox{0.8}{(a) TCN}}\medskip
\end{minipage}
\hfill
\begin{minipage}[b]{0.33\linewidth}
  \centering
  \centerline{\includegraphics[width=\linewidth]{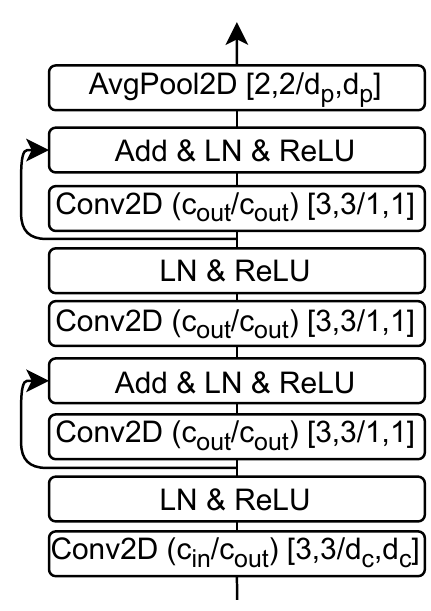}}
  \centerline{\scalebox{0.8}{(b) ResNetBlock}}\medskip
\end{minipage}
\hfill
\begin{minipage}[b]{0.36\linewidth}
  \centering
  \centerline{\includegraphics[width=\linewidth]{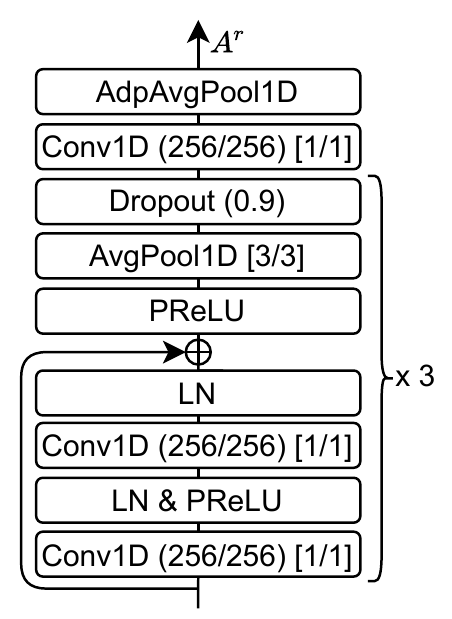}}
  \centerline{\scalebox{0.8}{(c) Speaker encoder}}\medskip
\end{minipage}
\caption{(a) The configuration of a temporal convolutional network, TCN [d]. $d$ is the stride in the $Conv1D^*$ layer, also referred to as the dilation factor. The symbol $\oplus$ represents element-wise addition. (b) The configuration of a ResNetBlock $(c_{in}/c_{out}) [d_c/d_p]$. Add refers to the element-wise addition. (c) The configuration of a speaker encoder.}
\label{fig:tcn&mts_resnet}
\end{figure}

The reentry model as shown in Fig.~\ref{fig:reentry} consists of a speech encoder, an attractor encoder, a speaker extractor, $R-1$ speaker encoders, and a speech decoder. The attractor encoder has the same configuration as the SLSyn network as detailed in Fig.~\ref{fig:mts}. The attractor encoder is followed by an adaptation layer of a stack of $5$ TCN [1], with the TCN configuration detailed in Fig.~\ref{fig:tcn&mts_resnet}(a). The $R-1$ speaker encoders share the same architecture, and the configuration is detailed in Fig.~\ref{fig:tcn&mts_resnet}(c). Each speaker encoder is preceded by a speech decoder and a speech encoder that have the shared weights with the speech encoder and speech decoder depicted in Fig.~\ref{fig:reentry}.

We use layer normalization instead of batch normalization for all models, such that a small batch size in the fine-tuning stage will not affect the normalization performance. The network parameters for the speech encoder, speaker encoders, speaker extractor, and speech decoder are selected following~\cite{wu2019time,pan2020muse}. In the speech encoder and speech decoder, the $L$ and $N$ are set to $40$ and $256$ respectively. In the speaker encoders, the $Dropout$ layer has a dropout probability of $0.9$. The speaker extractor has the $R$ and $B$ set to $4$ and $8$ respectively.

\subsection{Training of the \textit{reentry} model}
Both the SLSyn network and \textit{reentry} model are implemented with PyTorch\footnote{The data generation and the \textit{reentry} model codes are available at https://github.com/zexupan/reentry.}, and optimized by the Adam optimizer~\cite{kingma2015adam}, with an initial learning rate of $0.001$.

During the SLSyn network training, the learning rate decreases by $4\%$ every epoch, and the training stops when the best validation loss does not improve for 4 consecutive epochs. During the \textit{reentry} model training, the learning rate is halved when the best validation loss does not improve for 6 epochs consecutively, and the training stops when the best validation loss does not improve for 10 consecutive epochs. Both the SLSyn network and the \textit{reentry} model are trained on $2$ Tesla $32$GB V100 GPUs. The utterances are truncated to $6$ seconds to fit into the GPU memory during training, while the full utterance is evaluated at run-time inference.

The overall training is carried out in three stages. In the first stage, we pre-train the SLSyn network to form the attractor encoder. In the second stage, we train the \textit{reentry} model, which uses the pre-trained attractor encoder without updating its weights. In the last stage, we fine-tune the attractor encoder together with the \textit{reentry} model using the re-initialized optimizer. We will justify the 3-stage strategy through an ablation experiment.

\subsection{Baselines}
Time-domain speaker extraction algorithms usually outperform their frequency-domain counterparts. MuSE~\cite{pan2020muse} is a time-domain technique that reports the state-of-the-art performance in audio-visual speaker extraction, and is built on TDSE~\cite{wu2019time}. We implement the two models as the reference baselines.

\subsubsection{TDSE}
TDSE~\cite{wu2019time} has a network component and architecture similar to the \textit{reentry} model, except that the former has neither the speaker encoders for top-down auditory attention nor the speaker classification task, which is designed to train the speaker encoders. In addition, TDSE uses a pre-trained VSR network as the attractor encoder. The implementation of TDSE-O follows the original architecture in the paper. We also implement TDSE-I as an improved version for a fair comparison with the \textit{reentry} model. The improvements include, a) the TDSE-I replaces the batch normalization in TDSE-O with layer normalization, b) it concatenates the visual attractor to every stack of TCNs, just like what the \textit{reentry} model does, instead of the second stack only, and c) the attractor encoder is fine-tuned similar to the \textit{reentry} model.

\subsubsection{MuSE} 
Like TDSE, MuSE~\cite{pan2020muse} uses a VSR pre-trained network as the attractor encoder. We implement the original architecture, denoted as MuSE-O, and an improved version MuSE-I, with layer normalization and attractor encoder fine-tuning. MuSE-O takes the intermediate estimated speech embeddings $\hat{S}^r(t)$ and visual cue $V(t)$ as input to the speaker encoders, and applies a speaker encoder to the start of every TCN stack, while MuSE-I and \textit{reentry} model only take $\hat{S}^r(t)$ as input to the speaker encoders starting from the second TCN stack similarly to the \textit{reentry} model.

\subsubsection{Visual speech recognition}
\begin{figure}
  \centering
  \includegraphics[width=0.9\linewidth]{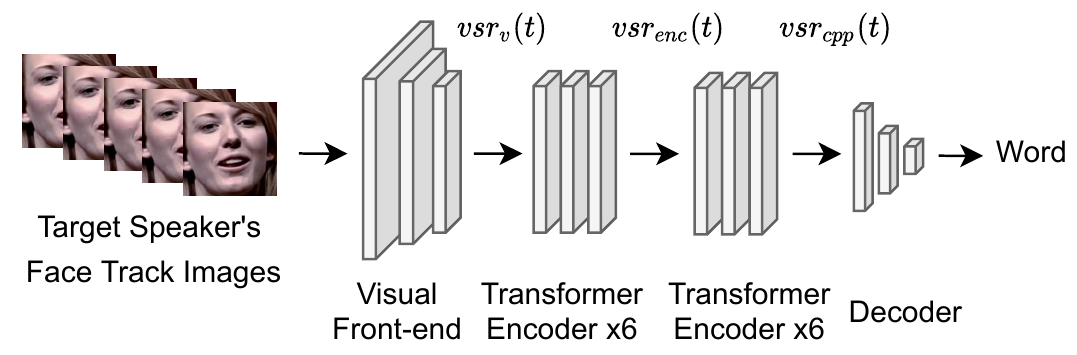}
  \caption{Three features at different abstraction level derived from a visual speech recognition model, namely $vsr_v(t)$, $vsr_{enc}(t)$, and $vsr_{cpp}(t)$.}
  \label{fig:vsr}
\end{figure}

TDSE and MuSE both use an attractor encoder pre-trained on the visual speech recognition task~\footnote{The visual speech recognition pre-trained models are taken from \url{https://github.com/lordmartian/deep_avsr}~\cite{Afouras18b}}. The attractor encoder generates $vsr_v(t)$, which is the output of the visual front-end of the VSR-LSTM model~\cite{stafylakis2017combining}. The visual front-end consists of a convolutional 3D and a ResNet 18 layer. In a visual speech recognition study~\cite{afouras2018deep}, this visual front-end is followed by $12$ transformer layers to form a VSR-TRAN model as illustrated in Fig.~\ref{fig:vsr}, where $vsr_{enc}(t)$ and $vsr_{cpp}(t)$ are the outputs of the $6$th and last layer of the transformer. $vsr_{cpp}(t)$ represents the character posterior probability at the frame level. We study the three features, namely $vsr_v(t)$ , $vsr_{enc}(t)$ and $vsr_{cpp}(t)$, which represent different level of abstraction of the visual cues in our experiments.

\subsection{Evaluation metrics}
We use the scale-invariant signal-to-noise ratio improvement (SI-SDRi); signal-to-noise ratio improvement (SDRi); perceptual evaluation of speech quality improvement (PESQi)~\cite{rix2001perceptual} and Short Term Objective Intelligibility improvement (STOIi) to evaluate the extracted speech (the higher the better for all metrics). The SI-SDRi and SDRi represent the improvements in the signal quality of the extracted speech qualities compared to the mixture speech. The PESQi and STOIi represent the improvements in the perceptual quality and intelligibility of the extracted speech compared to the mixture speech.

\begin{table*}
    \centering
    \caption{We report the evaluation of the pre-trained (PT) network VSR-TRAN in terms of word error rate (WER), and that of the SLSyn network in terms of binary classification accuracy (Acc.). We also evaluate the \textit{reentry} model on the VoxCeleb2-2mix dataset with different top-down attention attractors (\textit{Att}) derived from the pre-trained networks. The validation set (Val.) and test set SI-SDRi are reported with (w/) and without (w/o) the attractor encoder fine-tuning. The no. of network parameters (Params) is reported in million (m).}
    \addtolength{\tabcolsep}{-1pt}
    \begin{tabular}{c|c|c|c|c|c|c|c|c|c|c} 
      \toprule
        \multirow{2}*{PT network}                    &\multirow{2}*{PT Dataset}             &\multirow{2}*{Hours}  &WER    &Acc.   &\multirow{2}*{\textit{Att}}      &Val. SI-SDRi    &Test SI-SDRi    &Val. SI-SDRi    &Test SI-SDRi   &Params\\
        &&&(\%)&(\%)&&(dB) w/o&(dB) w/o&(dB) w/&(dB) w/&(m)\\
        \midrule
        \multirow{1}*{VSR-LSTM}
                                    &LRW~\cite{Chung16}     &160    &- &-              &$vsr_{v}$           &12.29      &11.88 &12.30       &11.91  &24.5\\
        \midrule
        \multirow{2}*{VSR-TRAN}
                                    &\multirow{2}*{LRS2~\cite{afouras2018deep}}
                                    &\multirow{2}*{200}
                                    &\multirow{2}*{61.8}
                                    &\multirow{2}*{-}  &$vsr_{enc}$                                         &12.07      &11.61  &12.12      &11.73      &75.1\\
                                    &&&&                                             &$vsr_{cpp}$           &11.25      &10.91  &11.51      &11.12  &130.1\\
        \midrule
        \multirow{4}*{SLSyn}
                                    &\multirow{4}*{VoxCeleb2-sync}
                                    &\multirow{4}*{2700}
                                    &\multirow{4}*{-}
                                    &\multirow{4}*{94.9}                          &$sync_{1}$             &12.17      &11.70  &12.27      &11.80  &\textbf{16.0}\\
                                    &&&&                                             &$sync_{2}$            &12.43      &12.14  &12.70      &12.34 &18.5\\
                                    &&&&                                             &$sync_{3}$            &\textbf{12.58} &\textbf{12.34}  &\textbf{13.01} &\textbf{12.60} &18.8\\
        
                                    &       &   & &         &a face image    &3.19 &2.92      &3.29 &3.00&18.8\\
        \midrule
        \multirow{2}*{SLSyn}
                                    &VoxCeleb2-sync-s       &270    &\multirow{2}*{-} &86.6         &\multirow{2}*{$sync_{3}$}          &12.27      &12.00 &12.87      &12.46  &18.8\\
                                    &VoxCeleb2-sync-c       &2700   & &96.8         &          &12.33      &11.91 &12.66      &12.26  &18.8\\
    
      \bottomrule
    \end{tabular}
    \addtolength{\tabcolsep}{1pt}
    \label{table:pretrain}
\end{table*}


\section{Experiments}
\label{sec:results}
\subsection{Visual speech recognition vs. speech-lip synchronization}
We first compare the two pre-training models, namely the visual speech recognition (VSR) network~\cite{Afouras18b} and our speech-lip synchronization (SLSyn) network, as reported in Table~\ref{table:pretrain}. As VSR is a challenging task with limited training data, the VSR-TRAN network has a high word error rate of 61.8\% on the LRS2 dataset.

For speech-lip synchronization detection, the accuracy is 86.6\% with 270 hours of training data (VoxCeleb2-sync-s). If 2,700 hours of data are used (VoxCeleb2-sync), the accuracy goes up to 94.9\%. If the dataset consists of clean speech only (VoxCeleb2-sync-c), the accuracy goes up to 96.8\%. The number (no.) of network parameters for the SLSyn network is less than the VSR network due to the simplicity of the task. In the VoxCeleb2-sync dataset, negative samples are target-shifted speech. To ensure the SLSyn network is not learning the shift instead of the synchronization, we conduct an additional negative sample test to classify the talking face track with only the interference speaker's speech. The SLSyn network achieves the classification accuracy of $96.0\%$, showing that the SLSyn network is learning the speech-lip synchronization information.

We evaluate the \textit{reentry} model on the VoxCeleb2-2mix dataset with different pre-trained attractor encoders in Table~\ref{table:pretrain}. In Section~\ref{sec:results}, we omit the subscript $(t)$ for the attention attractors \textit{Att} for brevity. With the VSR features, the lower level of abstraction is further away from the decision layer, and represents more the visual content than the phonetic information. We observe that the lower the feature abstraction level, the better the extracted signal quality, with $vsr_v$ having better results than $vsr_{enc}$ and $vsr_{cpp}$. The results suggest that the viseme-phoneme mapping objective, that optimizes $vsr_{cpp}$, is not the best for speaker extraction. 

With the SLSyn features, the higher level of abstraction is near the decision layer, and represents more the speech-lip synchronization/non-synchronization decision than the audio-visual content. We observe that the higher the abstraction level, the better the signal extraction quality produced, with $sync_{3}$ having better results than $sync_{2}$ and $sync_{1}$. When $sync_{3}$ is trained from the VoxCeleb2-sync-s dataset with a similar amount of data as VSR, it outperforms the best VSR feature $vsr_v$ by $0.55$dB in terms of SI-SDRi. The best synchronization feature $sync_{3}$ outperforms the best VSR feature $vsr_v$ by $0.69$dB in terms of SI-SDRi, showing the advantage of the speech-lip synchronization feature. Unless mentioned otherwise, the \textit{reentry} model reported in this paper uses $sync_3$ pre-trained from the VoxCeleb2-sync dataset for the speech-lip synchronization feature representation.

Among the synchronization features, the visual-only feature $sync_{1}$ lags behind audio-visual features $sync_{2}$ and $sync_{3}$ by about $0.5$dB in terms of SI-SDRi, showing the importance of the speech in the synchronization cue. When $sync_{3}$ is trained on the single-talker dataset (VoxCeleb2-sync-c), the SI-SDRi degrades by $0.34$dB, showing that the pre-training domain-mismatch affects the speaker extraction quality significantly.

The face-voice association can also be used as the top-down attention cue for speaker extraction~\cite{chung2020facefilter}. We would like to investigate if the attractor encoder in the \textit{reentry} model indeed benefits from the face-voice association or the speech-lip synchronization. We study the use of a single face image as the (\textit{Att}) in Table~\ref{table:pretrain}. During the \textit{reentry} model training, we first randomly pick a face image from the video sequence and repeat the same image to form an image sequence. We then use the image sequence in place of the video sequence as the top-down attention cue, and report an SI-SDRi of $3.00$dB, which is far behind that of $12.60$dB compared with $sync_{3}$. While the attractor encoder seems to take advantage of the face-voice association cue, it mainly benefits from the speech-lip synchronization cue, as most of SI-SDRi gain comes from the latter.

\subsection{Study of training strategy}

We perform an ablation study to justify the proposed 3-stage training strategy. In Table~\ref{table:training_stages}, Expt 1 represents a training strategy in which we train the entire \textit{reentry} model from scratch without any pre-training; Expt 2 represents a training strategy that uses the pre-trained SLSyn network to initialize the attractor encoder, then performs a full \textit{reentry} model training; Expt 3 represents a variant of Expt 2 that uses the pre-trained SLSyn network to initialize the attractor encoder, then performs a full reentry model training but freezing the attractor encoder; Expt 4 represents the proposed 3-stage training strategy, that is to initialize the attractor encoder with pre-trained SLSyn network, then fix the attractor encoder for model training, finally fine-tune the attractor encoder and the rest of the \textit{reentry} model together. As the $sync_{3}$ feature from the SLSyn network has the best representation, we adopt $sync_{3}$ to justify our training strategy for the \textit{reentry} model.

As shown in Table~\ref{table:training_stages}, the proposed 3-stage training strategy (Expt 4) provides an SI-SDRi improvement of $0.68$dB over that without pre-training (Expt 1), which suggests that the \textit{reentry} model benefits from the knowledge learned from the pre-training step. The proposed 3-stage training strategy (Expt 4) also outperforms that with a simple initialization (Expt 2) at retaining the transferred knowledge, with an SI-SDRi improvement of $0.58$dB. This could be because, during the first few epochs of speaker extractor training, the loss is high. In Expt 2, such high loss back-propagates and adversely affects the integrity of the attractor encoder. We observe in Expt 3, that by fixing the pre-trained attractor encoder, we gain $0.32$dB of SI-SDRi, and by fine-tuning the entire \textit{reentry} model resulting from Expt 3, we gain another $0.26$dB of SI-SDRi.

\begin{table}
\centering
    \caption{We study the training strategy for the \textit{reentry} model that involves the attractor encoder in four experiments. In all experiments, we train the \textit{reentry} model of the same architecture with different variants of the proposed training strategy on the VoxCeleb2-2mix dataset, and report the SI-SDRi for the validation and test sets. \textit{Init} denotes the use of pre-trained attractor encoder; \textit{Fix} denotes freezing the pre-trained attractor decoder; \textit{FT} denotes fine-tuning of the attractor encoder.}
    \addtolength{\tabcolsep}{-1pt}
    \begin{tabular}{c|c|c|c|c|c|c} 
      \toprule
                       \multirow{2}*{Expt}
                       &\multirow{2}*{\textit{Att}}  & \multirow{2}*{\textit{Init}}         & \multirow{2}*{\textit{Fix}}   & \multirow{2}*{\textit{FT}}   & Val. SI-SDRi    &Test SI-SDRi\\
                       &  &          &    &     &  (dB)    &(dB) \\
        \midrule
                             1 &\multirow{4}*{$sync_{3}$} &\xmark             &N/A            &N/A            &12.27  &11.92\\
                             2 &&\cmark             &\xmark         &N/A            &12.42  &12.02\\
                             3 &&\cmark             &\cmark         &\xmark         &12.58  &12.34\\
                             4 &&\cmark             &\cmark         &\cmark         &\textbf{13.01} & \textbf{12.60}\\
        
      \bottomrule
    \end{tabular}
    \addtolength{\tabcolsep}{1pt}
    \label{table:training_stages}
\end{table}

\begin{table}
    \centering
    \caption{A comparison among the \textit{reentry} model, and its variants on the VoxCeleb2-2mix dataset. $\gamma$ is the scaling factor in the loss function. $R$ is the no. of TCN stacks in the speaker extractor pipeline. `Shared Weights' denotes a network architecture where the weights in $R-1$ speaker encoders are shared. `with $A^r$' denotes the use of the self-enrolled speaker embedding $A^r$ as part of the top-down attention cue.}
    \addtolength{\tabcolsep}{-1pt}
    \begin{tabular}{c|c|c|c|c|c|c} 
       \toprule
        \multirow{2}*{System}               &\multirow{2}*{$\gamma$}   &\multirow{2}*{R}  &Shared    &With      &Val. SI-SDRi &Test SI-SDRi\\ 
        &&&Weights&$A^r$&(dB)&(dB) \\
        \midrule
        1   &0.5            &4&\multirow{7}*{\xmark}&\multirow{7}*{\cmark}&12.51  &12.03\\
        2   &0.1            &4&&&12.92  &12.51\\
        3   &0.05           &4&&&12.89  &12.45\\
        4   &0.01           &4&&&12.99  &12.52\\
        5   &0.005          &4&&&\textbf{13.01} &\textbf{12.60}\\
        6   &0.001          &4&&&12.90  &12.51\\
        7   &0              &4&&&12.68  &12.27\\
        \midrule
        8   &0.005          &3&\multirow{2}*{\xmark}&\multirow{2}*{\cmark}&11.87 &11.90\\
        9   &0.005          &2&&&11.07 &11.13\\
        \midrule

        10     &N/A &\multirow{3}*{4}  &N/A        &N/A&12.64  &12.30\\
        11     &0.005 &  &\xmark  &\xmark        &12.77 &12.39\\
        12     &0.005 &  &\cmark  &\cmark        &12.83 &12.46\\
        \midrule
        13 
                            &0.1            &\multirow{3}*{4}   &\multirow{3}*{\xmark}   &\multirow{3}*{\cmark}&\textbf{13.40}  &13.01\\
        14  &0.005          &&&&13.38 &\textbf{13.02}\\
        15 &0              &&&&13.34  &12.95\\
        \midrule
        16        &N/A            &4  &N/A &N/A  &13.15  &12.72\\
       \bottomrule
    \end{tabular}
    \addtolength{\tabcolsep}{1pt}
    \label{table:se}
\end{table}

\subsection{Study of speaker encoders} 

We study the self-enrolled speaker encoders to justify their role through experiments reported in Table~\ref{table:se}.

System 10 has the same architecture as systems 1-7, except that the former has neither the speaker encoders nor the speaker classification task, which is designed to train the speaker encoders. In the speaker extractor in Fig.~\ref{fig:reentry} , the interlaced architecture between the speaker encoders and the TCN stacks is designed to estimate the mask $M(t)$. With system 10, we would like to examine the contribution of the speaker encoders.

For systems 1-7, we vary the scaling factor $\gamma$, which weights the contributions between speaker classification and speaker extraction during the multi-task learning. With $\gamma=0$ (system 7), we ignore the speaker classification task during the training of speaker encoders and obtain an SI-SDRi of $12.27$dB, which is similar to that of system 10. The results suggest that the $R-1$ speaker encoders are contributing only when they are trained under speaker classification supervision in multi-task learning. When $\gamma=0.005$ (system 5), the \textit{reentry} model achieves the best SI-SDRi of $12.60$dB, which represents $0.33$dB improvement over that for $\gamma=0$ (system 7). The results clearly suggest that the speaker encoders benefit from multi-task learning.

We also present systems 13-15 as an upper-bound model, where the speaker encoders take the target speaker's clean speech as the input instead of the intermediate speech, during both training and evaluation. System 14 delivers the best SI-SDRi of $13.02$dB. The proposed \textit{reentry} model (system 5) lags only $0.42$dB behind the upper-bound (system 14). 

Furthermore, we build system 16, which uses a speaker embedding extracted from the target speaker's clean speech with a pre-trained speaker verification model. During both training and evaluation, the speaker extractor takes such speaker embedding as input instead of speaker embedding estimated on the fly. Despite having clean speech as input, system 16 only attains an SI-SDRi of $12.72$dB. This suggests that the independently trained speaker encoders don't work the best for speaker extraction, thus once again justifies the proposed joint training between the speaker encoders and the speaker extraction network. 

\subsection{Study of speaker extractor}
We consider that the speaker encoders and speaker classification task may jointly contribute to the speaker extraction in two ways. First, $A^r$ provides the top-down attention cue for speaker extraction. Second, the speaker classification task supervises the training of speaker encoders and TCN stacks to ensure the output speech carries the target speaker identity. The question is in what way the speaker encoders with speaker classification task contribute to the extracted speech quality.

To answer the question, we implement system 11 as shown in Table~\ref{table:se}, in which we alter the standard \textit{reentry} model (system 5) by keeping the speaker encoders during multi-task learning, but removing the top-down attention cue $A^r$ from the interlace. During training, the gradients of the speaker classification loss back-propagate through the speaker encoders to update the TCN stacks, therefore, ensuring the speaker identity of the estimated speech. However, at run-time, only the TCN stacks pipeline participates in the mask estimation without $A^r$. 

The difference between system 11 and system 10 is that the former involves the speaker encoders, as well the speaker classification loss, during training, but the latter does not. In Table~\ref{table:se}, we observe that system 11 has improved SI-SDRi over system 10 from $12.30$dB to $12.39$dB, which shows the speaker classification task, without explicitly using $A^r$, offers a small gain of $0.09$dB. However, when $A^r$ is used as the attractor along with $V(t)$ in the \textit{reentry} model (system 5), we obtain a higher SI-SDRi of $12.60$dB. The results suggest that the use of self-enrolled $A^r$ as a top-down attention cue dominates the contribution from the speaker encoders and speaker classification task.

The speaker extractor in the \textit{reentry} model consists of $R$ TCN stacks that interlace with $R-1$ speaker encoders. We conduct further experiments to show the interlaced architecture supports a progressively refined speaker self-enrollment process, and report in Table~\ref{table:se}. We vary the number of TCN stacks $R$ from 2 to 4 in systems 9, 8, and 5, thus the speaker encoders $R-1$ from 1 to 3, in the speaker extractor. It is seen that, as $R$ increases, the SI-SDRi of \textit{reentry} increases from $11.13$dB to $12.60$dB at $\gamma=0.005$, showing that the progressive refinement architecture outperforms a single-stage extraction. 

We finally analyze the weight sharing among the speaker encoders. System 12 shares the weights across $R-1$ speaker encoders, while system 5 has individual weights for each speaker encoder. The results suggest that the individual weights scheme outperforms the shared weights counterpart by $0.14$dB in terms of SI-SDRi, which confirms our intuition to adopt individual weights in the \textit{reentry} model.

\begin{table*}
    \centering
    \caption{Performance of the \textit{reentry} model and baselines on the test set of simulated VoxCeleb2 dataset mixtures.}
    \begin{tabular}{c|c|c|c|c|c|c|c|c} 
       \toprule
        Dataset Mixtures &Task            &Model                &\textit{Att}     &SI-SDRi (dB)        &SDRi (dB)   &PESQi          &STOIi      &Params (m)    \\ 
        \midrule
        \multirow{9}*{VoxCeleb2-2mix}
        &\multirow{2}*{Speech separation}
                        &Conv-TasNet~\cite{luo2019conv}          &N/A            &10.94          &11.27       &0.835           &0.230           &8.7\\
        &               &DPRNN~\cite{luo2020dual}               &N/A            &11.57          &11.88       &0.902           &0.235           &2.6\\
        \cmidrule{2-9}
        &\multirow{5}*{Speaker extraction}
                        &TDSE-O~\cite{wu2019time}           &\multirow{4}*{$vsr_{v}$}       &10.64          &10.93  &0.823      &0.226      &20.2\\
        &                &\multirow{1}*{TDSE-I}             &      &11.62          &11.96       &0.958           &0.240           &20.1\\
        &               &MuSE-O~\cite{pan2020muse}          &               &11.67          &12.04  &0.969      &0.241      &25.0\\
        &                &MuSE-I                            &      &11.91          &12.22       &1.007           &0.245           &24.5\\
        \cmidrule{3-9}
        &                &\textit{reentry}                                  &$sync_{3}$   &\textbf{12.60} &\textbf{12.92}       &\textbf{1.103}           &\textbf{0.256}           &18.8\\
        \midrule
        \multirow{9}*{VoxCeleb2-3mix}
        &\multirow{2}*{Speech separation}
                        &Conv-TasNet~\cite{luo2019conv}              &N/A            &9.31          &9.74       &0.264           &0.230           &8.8\\
        &                &DPRNN~\cite{luo2020dual}                  &N/A            &9.83          &10.23       &0.283           &0.239           &2.7\\
        \cmidrule{2-9}
        &\multirow{5}*{Speaker extraction}
                        &TDSE-O~\cite{wu2019time}               &\multirow{4}*{$vsr_{v}$}       &9.78          &10.21  &0.360      &0.262      &20.2\\
        &               &\multirow{1}*{TDSE-I}                  &      &11.54          &12.04       &0.500           &0.303           &20.1\\
        &               &MuSE-O~\cite{pan2020muse}              &               &11.64          &12.18  &0.513      &0.306      &25.0\\
        &               &MuSE-I                                 &      &12.21          &12.68       &0.559           &0.318           &24.5\\
        \cmidrule{3-9}
        &               &\textit{reentry}                                       &$sync_{3}$   &\textbf{12.63} &\textbf{13.08}       &\textbf{0.613}           &\textbf{0.327}           &18.8\\
       \bottomrule
    \end{tabular}
    \label{table:benchmark}
\end{table*}

\subsection{Benchmark against baselines}
We now compare the \textit{reentry} model with two recent models TDSE and MuSE on mixture speech of two-speaker (VoxCeleb2-2mix) and three-speaker (VoxCeleb2-3mix) datasets as summarized in Table~\ref{table:benchmark}. We observe that the improved versions of TDSE and MuSE, namely TDSE-I and MuSE-I outperform their original model, i.e., TDSE-O~\cite{wu2019time} and MuSE-O~\cite{pan2020muse}, thus serving as competitive baselines. Since MuSE-I employs the same speaker encoders architecture and multi-task learning framework as the \textit{reentry} model, we used the $\gamma=0.005$ for MuSE-I training as it proves to learn the best speaker representation according to Table~\ref{table:se}. It is seen that the \textit{reentry} model outperforms all the baselines in terms of SI-SDRi, SDRi, PESQi, and STOIi on the VoxCeleb2-2mix and VoxCeleb2-3mix datasets.

We also re-implement two speech separation models, namely Conv-TasNet~\cite{luo2019conv} and DPRNN~\cite{luo2020dual} following the original paper, and report their results on the VoxCeleb2-2mix and VoxCeleb2-3mix datasets for comparison. Since Conv-TasNet and DPRNN are audio-only speech separation algorithms, they don't use the visual information from the VoxCeleb2-2mix dataset mixture. Hence we expect that the audio-visual speaker extraction outperforms the audio-only models.

\begin{figure*}
\begin{minipage}[t]{.32\linewidth}
  \centering
  \centerline{\includegraphics[width=6.6cm, height=4.4cm]{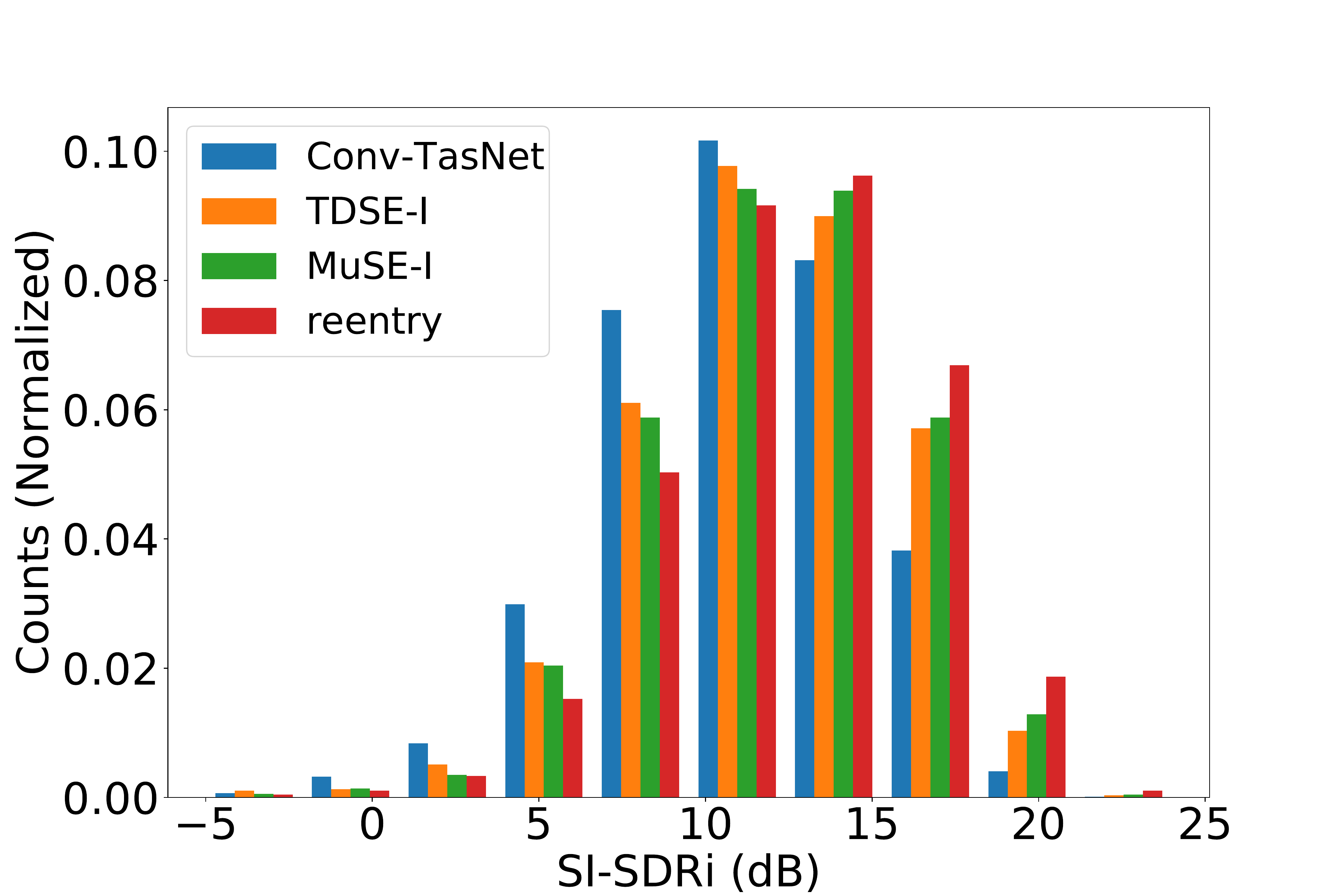}}
  \caption{The histogram of SI-SDRi by Conv-TasNet, TDSE-I, MuSE-I and \textit{reentry} model on VoxCeleb2-2mix.}\medskip
  \label{fig:count_sisdri}
\end{minipage}
\hfill
\begin{minipage}[t]{0.32\linewidth}
  \centering
  \centerline{\includegraphics[width=6.6cm, height=4.4cm]{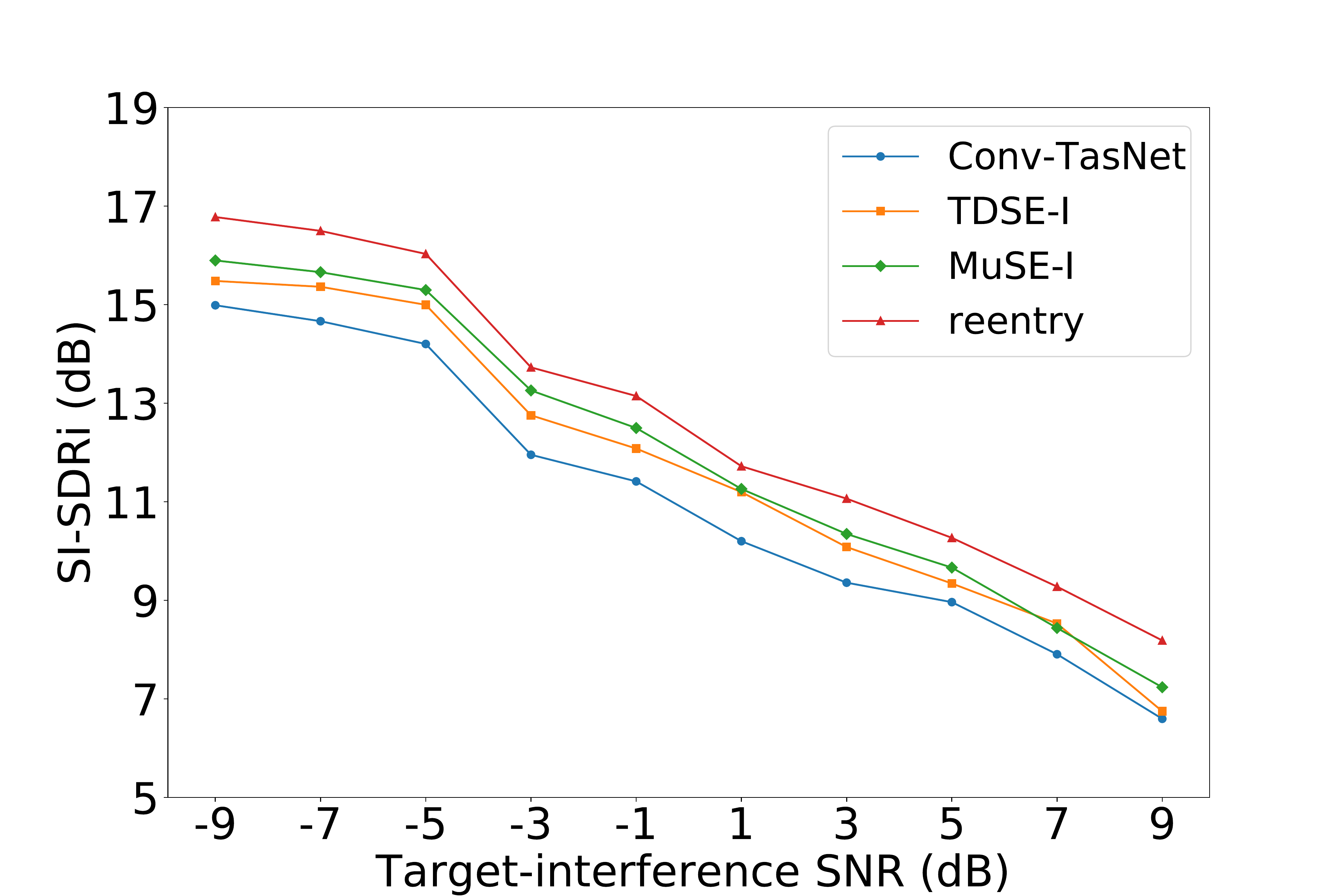}}
  \caption{The average SI-SDRi by Conv-TasNet, TDSE-I, MuSE-I and \textit{reentry} model on VoxCeleb2-2mix as a function of the target-interference SNR. }\medskip
  \label{fig:sisdr_snr}
\end{minipage}
\hfill
\begin{minipage}[t]{0.32\linewidth}
  \centering
  \centerline{\includegraphics[width=6.6cm, height=4.4cm]{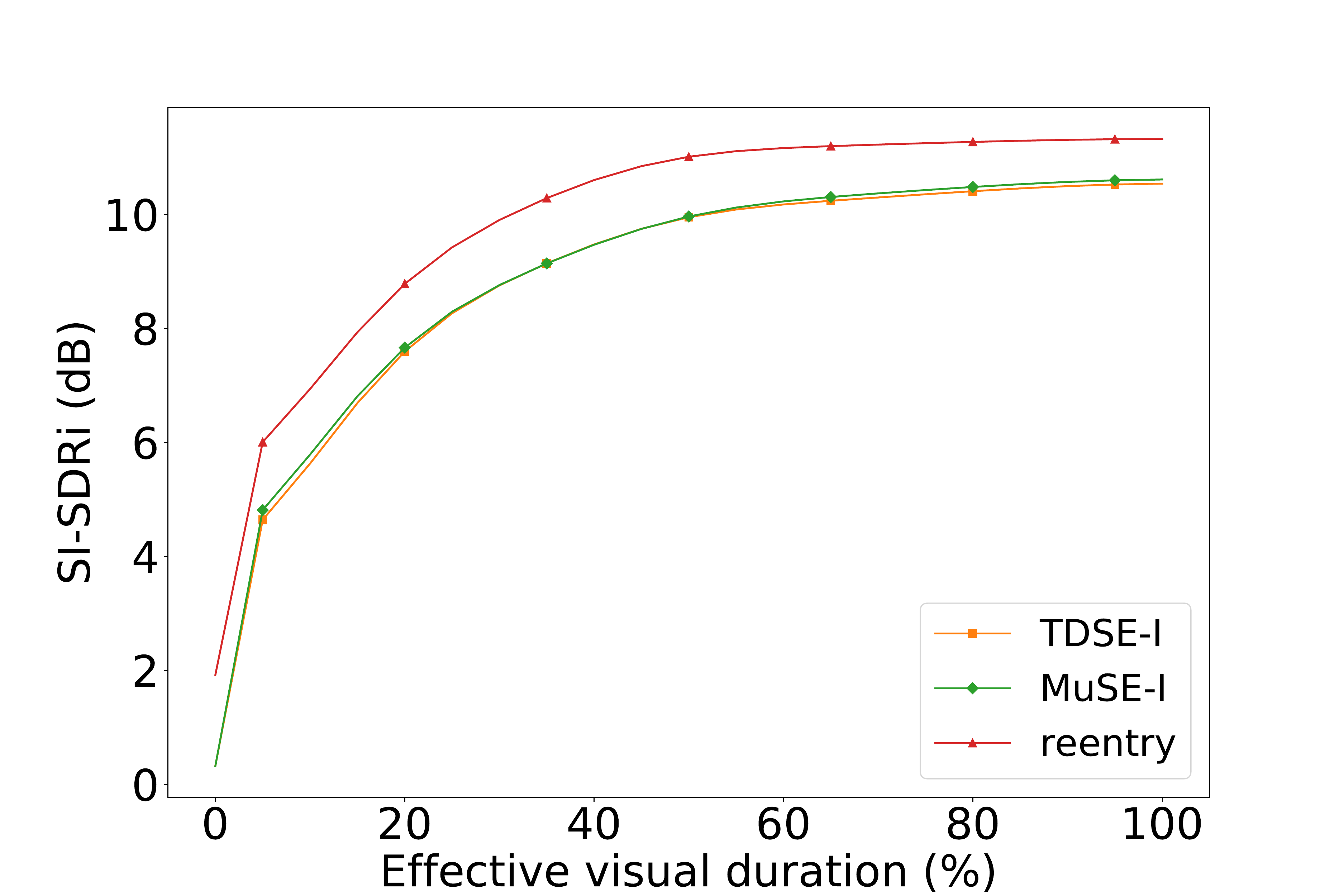}}
  \caption{The average SI-SDRi by TDSE-I, MuSE-I and \textit{reentry} model on VoxCeleb2-2mix-occl as a function of effective visual duration.}
  \medskip
  \label{fig:occl}
\end{minipage}
\end{figure*}

In Fig.~\ref{fig:count_sisdri}, we show the histogram of SI-SDRi for the VoxCeleb2-2mix test set samples by the Conv-TasNet, TDSE-I, MuSE-I, and the \textit{reentry} model. It shows that the \textit{reentry} model has more extracted speech samples with higher SI-SDRi gain than the baselines. In Fig.~\ref{fig:sisdr_snr}, we show the distribution of SI-SDRi gain for mixtures with various target-interference SNR. The target-interference SNR is defined as the energy contrast between the target speaker and the interference speaker in terms of SNR. It is seen that the \textit{reentry} model consistently achieves a higher SI-SDRi than the baselines. As the input mixture becomes less noisy (high SNR), the SI-SDRi gain becomes smaller.

\begin{table}
    \centering
    \caption{The results on the VoxCeleb2-2mix-occl test set for models that trained with and without occlusion data.}
    \addtolength{\tabcolsep}{-1pt}
    \begin{tabular}{c|c|c|c|c|c} 
      \toprule
        Model       &Train data                     &SI-SDRi   &SDRi        &PESQi &STOIi \\
        \midrule
        TDSE-I      &\multirow{3}*{VoxCeleb2-2mix}  &4.98       &6.18       &0.671  &0.091\\
        MuSE-I      &                               &4.02       &5.40           &0.625  &0.066\\
        \textit{reentry}        &                   &\textbf{6.68}  &\textbf{7.70}      &\textbf{0.801} &\textbf{0.125}\\
        \midrule
        TDSE-I      &\multirow{3}*{VoxCeleb2-2mix-occl}&9.21           &9.82    &0.856 &0.188\\
        MuSE-I    &                                 &9.29           &9.87       &0.875 &0.188\\
        \textit{reentry}        &                   &\textbf{10.27} &\textbf{10.74}       &\textbf{0.909} &\textbf{0.209}\\
      \bottomrule
    \end{tabular}
    \label{table:occl}
    \addtolength{\tabcolsep}{1pt}
\end{table}

\subsection{Speaker extraction in face of visual occlusion}
As the proposed \textit{reentry} model relies on the speech-lip synchronization cue, any visual occlusion will adversely impact the performance. We evaluate three systems, namely the \textit{reentry} model,  TDSE-I, and MuSE-I to observe how systems perform in face of visual occlusions. 

We create a two-speaker visual occlusion dataset, named the VoxCeleb2-2mix-occl, based on the VoxCeleb2-2mix dataset, but blackout the visual signals randomly and for a random duration while keeping the audio signals intact. This simulates the scenarios where the face-tracking algorithm fails to detect the presence of the target speaker.

In Table~\ref{table:occl}, it is observed that none of the models generalizes well if not trained on occlusion data. Nonetheless, the \textit{reentry} model outperforms the best baselines for all evaluation metrics. When trained on occlusion data, all models improve, with the \textit{reentry} model still showing the best performance. 

We analyze the performance with different effective visual duration after the visual occlusion in Fig.~\ref{fig:occl}. The effective visual duration is defined as the percentage of the duration of the unoccluded visual reference to that of the entire utterance. We plot the SI-SDRi of the \textit{reentry} model and baselines with different effective visual duration. It is seen that when the visual information is present for about 10\%, the \textit{reentry} model is able to achieve near $7$dB SI-SDRi. When the visual information is present for about 50\%, the performance of the \textit{reentry} model starts to saturate, the model is able to extract the speech segments where the visual information is occluded according to the prosody of the extracted speech when the visual information is present. 

\begin{table}
    \centering
    \caption{Cross-datasets evaluations for models that are trained on the VoxCeleb2-2mix dataset and tested on other datasets.}
    \label{table:result_cross_domain}
    \addtolength{\tabcolsep}{-1pt}
    \begin{tabular}{c|c|c|c|c|c|c} 
       \toprule
        Domain   &Dataset         & Model               &SI-SDRi &SDRi  &PESQi &STOIi\\        
       \midrule
       \multirow{6}*{Wild}    &\multirow{3}*{LRS2}        &TDSE-I        &11.61 &12.05 &0.874 &0.234\\
                                                         &&MuSE-I        &12.21 &12.60 &0.937 &0.245\\
                                                        &&\textit{reentry}  &\textbf{12.66} &\textbf{13.08} &\textbf{1.036} &\textbf{0.250}\\ \cmidrule{2-7}
                            &\multirow{3}*{LRS3}        &TDSE-I       &13.39 &13.71 &1.102 &0.241\\
                                                        &&MuSE-I      &13.69 &13.98 &1.142 &0.245\\
                                                        &&\textit{reentry}    &\textbf{14.24} &\textbf{14.54} &\textbf{1.244}&\textbf{0.251} \\
       \midrule
       \multirow{6}*{Studio}   & \multirow{3}*{Grid}       &TDSE-I     &9.33 &11.51 &0.822 &0.077\\
                                                            &&MuSE-I      &\textbf{10.18} &12.06 &\textbf{0.931} &0.090\\
                                                        &&\textit{reentry}     &9.83  &\textbf{12.09} &0.893 &\textbf{0.098} \\ \cmidrule{2-7}
                            &\multirow{3}*{TCD-TIMIT}   &TDSE-I        &13.70 &14.22  &0.989 &0.174\\
                                                        &&MuSE-I     &14.36 &14.87 &1.067 &0.183 \\
                                                        &&\textit{reentry}   &\textbf{14.96} &\textbf{15.55}  &\textbf{1.118} &\textbf{0.188} \\
       \bottomrule
    \end{tabular}
    \addtolength{\tabcolsep}{1pt}
\end{table}

\subsection{Cross-dataset evaluation}
We are interested in how the \textit{reentry} model trained on the VoxCeleb2 dataset performs on other datasets. We evaluate the \textit{reentry} model with two competing models on Grid~\cite{cooke2006audio}, TCD-TIMIT~\cite{harte2015tcd}, LRS2~\cite{afouras2018deep}, and LRS3~\cite{afouras2018lrs3} datasets. Grid and TCD-TIMIT are `studio' videos while LRS2 and LRS3 are `wild' videos from BBC and TED. These datasets are pre-processed with face detection and tracking algorithms same as the VoxCeleb2 dataset to minimize the visual mismatch. We generate 3000 utterances to form a test set for each of the above datasets, following the VoxCeleb2-2mix protocol.

As shown in Table \ref{table:result_cross_domain}, on the LRS2 and LRS3 datasets, the \textit{reentry} model shows higher SI-SDRi, SDRi, PESQi, and STOIi, which are consistent with the improvement on the VoxCeleb2 dataset. On the Grid and TCD-TIMIT datasets, which belong to another domain, the \textit{reentry} model still shows a relative improvement over the competing models, with the exception that MuSE-I outperforms the \textit{reentry} model on the Grid dataset for SI-SDRi and PESQi.

\section{Conclusion}
\label{sec:conclusion}
We propose an end-to-end audio-visual speaker extraction network to emulate human selective auditory attention in the cocktail party. The study is motivated by the \textit{reentry} theory by Edelman which shows the cross-modal temporal integration in human auditory attention. The proposed \textit{reentry} model is particularly useful in situations where a pre-enrolled face or speech reference is not available. In summary, the \textit{reentry} model presents a step forward in solving the cocktail party problem using a computational model. It will potentially enable applications, such as intelligent hearing aids, or downstream tasks such as speech recognition and speaker verification.


%





\ifCLASSOPTIONcaptionsoff
  \newpage
\fi





\bibliographystyle{IEEEtran}
\bibliography{IEEEabrv,Bibliography}

%






\begin{IEEEbiography}[{\includegraphics[width=1in,height=1.25in]{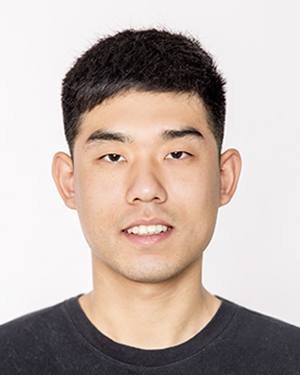}}]{Zexu Pan} received the B.Eng. degree in electrical and electronic engineering from Nanyang Technological University, Singapore, in 2019. He is currently a Ph.D. candidate at the Institute of Data Science, National University of Singapore, Singapore. His research interests include speech extraction, speech enhancement, multi-talker robust automatic speech recognition, multi-modality representation learning.
\end{IEEEbiography}

\begin{IEEEbiography}[{\includegraphics[width=1in,height=1.25in]{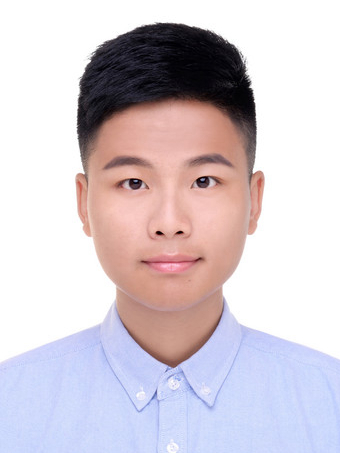}}]{Ruijie Tao} received the B.Eng. degree from Soochow University, China, in 2018. He received the M.Sc. degree from National University of Singapore, Singapore, in 2019. He is currently a Ph.D. candidate at National University of Singapore, Singapore. His research interests include speaker recognition, speech extraction, audio-visual representation learning.
\end{IEEEbiography}

\begin{IEEEbiography}[{\includegraphics[width=1in,height=1.25in]{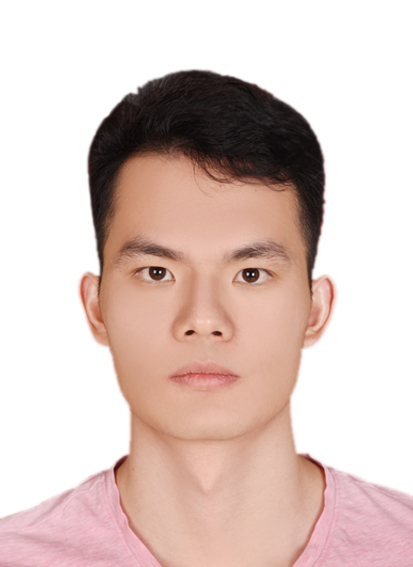}}]{Chenglin Xu}
(Member, IEEE) received the B.Eng. and M.Sc. degrees from Northwestern Polytechnical University, China, in 2012 and 2015, respectively. He obtained his Ph.D. degree from Nanyang Technological University, Singapore, in 2020. Prior to join Kuaishou Technology, he was a Research Fellow of HLT lab in Department of Electrical and Computer Engineering, National University of Singapore. His research interests include speech extraction, speech enhancement, source separation, multi-talker speaker verification and robust speech recognition.
\end{IEEEbiography}

\begin{IEEEbiography}[{\includegraphics[width=1in,height=1.25in]{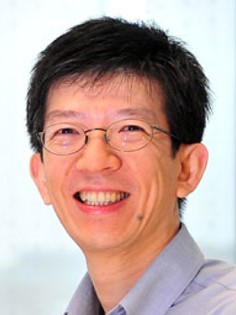}}]{Haizhou Li} 
(Fellow, IEEE) received the B.Sc., M.Sc., and Ph.D. degree in electrical and electronic engineering from South China University of Technology, Guangzhou, China in 1984, 1987, and 1990 respectively. He is currently a Presidential Chair Professor with the School of Data Science, The Chinese University of Hong Kong, Shenzhen, China. He is also an Adjunct Professor with the Department of Electrical and Computer Engineering, National University of Singapore (NUS). Prior to that, he taught in the University of Hong Kong (1988-1990) and South China University of Technology (1990-1994). He was a Visiting Professor at CRIN in France (1994-1995), Research Manager at the Apple-ISS Research Centre (1996-1998), Research Director in Lernout \& Hauspie Asia Pacific (1999-2001), Vice President in InfoTalk Corp. Ltd. (2001-2003), and the Principal Scientist and Department Head of Human Language Technology in the Institute for Infocomm Research, Singapore (2003-2016). His research interests include automatic speech recognition, speaker and language recognition, natural language processing.
Dr Li was the Editor-in-Chief of IEEE/ACM Transactions on Audio, Speech and Language Processing (2015-2018), a Member of the Editorial Board of Computer Speech and Language since 2012, an elected Member of IEEE Speech and Language Processing Technical Committee (2013-2015), the President of the International Speech Communication Association (2015-2017), the President of Asia Pacific Signal and Information Processing Association (2015-2016), and the President of Asian Federation of Natural Language Processing (2017-2018). He was the General Chair of ACL 2012, INTERSPEECH 2014, ASRU 2019 and ICASSP 2022.
Dr Li is a Fellow of the ISCA, and a Fellow of the Academy of Engineering Singapore. He was the recipient of the National Infocomm Award 2002, and the President’s Technology Award 2013 in Singapore. He was named one of the two Nokia Visiting Professors in 2009 by the Nokia Foundation, and U Bremen Excellence Chair Professor in 2019.

\end{IEEEbiography}


\vfill


\end{document}